\documentclass[aps,12pt,final,notitlepage,oneside,onecolumn,nobibnotes,nofootinbib,superscriptaddress,noshowpacs,centertags]{revtex4}

\RequirePackage[cp1251]{inputenc}

\usepackage{amsmath,amssymb,epsfig,placeins,subfigure,wrapfig,indentfirst}

\begin{document}

\title{On stability of a new model of wormhole}

\author{Igor Novikov}
\affiliation{Astro Space Center, Lebedev Physical Institute,
Russian Academy of Sciences, Moscow, Russia} \affiliation{Niels
Bohr Institute, Copenhagen, Denmark} \affiliation{Kurchatov
Institute,  Moscow, Russia}
\author{Alexander Shatskiy\footnote{shatskiy@asc.rssi.ru}}
\affiliation{Astro Space Center, Lebedev Physical Institute,
Russian Academy of Sciences, Moscow, Russia}
\date{\today}

\begin{abstract}
{\bf Abstract}\\

We investigate the stability of the wormhole of the
Mors-Thorne-Ellis-Bronnikov type. In our model the matter inside
it consists of a radial magnetic field and an ideal phantom-fluid.
Properties of the matter are described in section~\ref{s1},
\ref{s2} and \ref{s3} of this paper. We consider spherical
perturbations only and find examples of the stable wormholes
against these perturbations.

\end{abstract}

\maketitle

\section{INTRODUCTION}
\label{s1}

The problem of the stability of the wormholes has been discussed
in many papers (see~\cite{Armendariz-Picon2002-1,
Armendariz-Picon2002-2, Shinkai2002, Gonzalez2008-1,
Gonzalez2008-2, Doroshkevich2008-2, Doroshkevich2009-1,
Novikov2009-1, Novikov2009-2, Sarbach2010}).

The simplest model of a wormhole~\cite{Morris1988-2,
Bronnikov1973, Ellis1973} (phantom scalar field with the negative
kinetic term) proved to be unstable (in spite of erroneous
papers~\cite{Armendariz-Picon2002-1, Armendariz-Picon2002-2})  and
in accordance with more recent (independent) correct
research~\cite{Shinkai2002, Gonzalez2008-1, Gonzalez2008-2,
Doroshkevich2008-2, Doroshkevich2009-1, Bronnikov2011}.

A more complex model of a wormhole (with the same
metric~\cite{Doroshkevich2008-2, Doroshkevich2009-1} --- radial
magnetic field and the phantom dust with the negative energy
density) is almost stable under all types of the spherical
perturbation (except for the longitudinal radial motion of the
dust on the inertia). The growth of unstable type is sufficiently
slow: it increases proportionally to the time~\cite{Novikov2009-1,
Novikov2009-2}. Therefore, the authors of this paper it has been
suggested that the unstable type can be easily crushed by the
introduction into the model additional parameters. In the
paper~\cite{Visser2011} it was found a stable thin-shell
traversable wormholes model, with a thin shell of phantom matter
on its throat (by using the cut-and-paste procedure).

All of these studies provided hope for the existence and finding
fully stable solution (under all spherical perturbation types).

In this paper we investigate the stability (under the spherical
perturbations) the similar model of the
Mors-Thorne-Ellis-Bronnikov type wormhole (MTEB), see
also~\cite{Novikov2012}.

The matter of our model of the wormhole (WH) consists of the
radial monopole magnetic field (with topological charge $q$) and
the ideal phantom fluid. In the stationary case (respect to which
will be a study on stability) the energy density $\varepsilon$ of
this fluid is negative and equal to twice of the absolute value of
the energy density of the magnetic field. The pressure $p$ of the
fluid in a stationary case is equal to zero (phantom dust). In the
case of deviation of the energy density $\varepsilon$ from its
stationary value --- the pressure $p$ is proportional to this
deviation ${f\propto\varepsilon -\varepsilon_0}$. As already
mentioned, without pressure, this model is linearly unstable over
time~\cite{Novikov2009-1, Novikov2009-2}.

\section{Equations of the model}
\label{s2}

It is convenient to choose the metric tensor in the spherically
symmetric case like the following\footnote{Units are chosen as:
${c=1}$ and ${G=1}$ (the speed of light and gravitational
constant).}:
\begin{equation}
ds^2=e^\nu dt^2 – e^\lambda dx^2 – e^\alpha (d\theta^2 +
\sin^2\theta\, d\varphi^2) . \label{1-1}\end{equation} Here
$e^\alpha =r^2$, where $4\pi r^2$ - area of a sphere around the
center of the system, $r$, $\nu$ and $\lambda$ are functions $x$
and $t$.

The Einstein equations corresponding to the metric (\ref{1-1}), in
the comoving reference system~\cite{Landau1} can be written
as\footnote{The derivation of these equations can be found, for
example in~\cite{Landau1} (task 5 in \S 100), prime denotes the
derivative with respect to $x$, and dot - with respect to $t$.}:
\begin{eqnarray}
8\pi\varepsilon +q^2 e^{-2\alpha} = -e^{-\lambda}\left[ \alpha''
+{3\over 4}(\alpha')^2 - {1\over 2}\alpha’\lambda' \right] +
e^{-\nu}
\left[{1\over 2}\dot\alpha\dot\lambda +{1\over 4}(\dot\alpha)^2 \right] +e^{-\alpha}\, , \label{1-2}\\
8\pi p -q^2 e^{-2\alpha} = -e^{-\nu}\left[ \ddot\alpha + {3\over
4}(\dot\alpha)^2 - {1\over 2}\dot\alpha\dot\nu \right] +
e^{-\lambda}
\left[{1\over 2}\alpha'\nu' +{1\over 4}(\alpha')^2 \right]– e^{-\alpha}\, , \label{1-3}\\
8\pi p +q^2 e^{-2\alpha} = e^{-\lambda}\left[{1\over 2}\nu'' +
{1\over 4}(\nu')^2 + {1\over 2}\alpha'' +{1\over 4}(\alpha')^2 –
{1\over 4}\alpha'\lambda' - {1\over 4}\nu'\lambda' + {1\over 4}\alpha'\nu' \right] +\nonumber\\
+ e^{-\nu}\left[{1\over 4}\dot\lambda\dot\nu +{1\over
4}\dot\alpha\dot\nu – {1\over 4}\dot\lambda\dot\alpha - {1\over
2}\ddot\lambda - {1\over 4}(\dot\lambda)^2
 -{1\over 2}\ddot\alpha - {1\over 4}(\dot\alpha)^2 \right]\, , \label{1-4} \\
2\dot\alpha' +\dot\alpha\alpha' -\dot\lambda\alpha' -
\nu'\dot\alpha =0\, . \label{1-5}
\end{eqnarray}

The stationary wormhole MTEB metric tensor is defined by
\begin{equation}
ds^2=dt^2 – dx^2 – (q^2+x^2)\cdot (d\theta^2 + \sin^2\theta\,
d\varphi^2) . \label{1-6}\end{equation} This WH has the
energy-momentum tensor which corresponds to a mixture of monopole
electric (or magnetic) field and the phantom dust:
\begin{equation}
8\pi T^n_m= diag \left\{ \frac{+q^2}{(q^2+x^2)^2} ,\,
\frac{+q^2}{(q^2+x^2)^2} ,\, \frac{-q^2}{(q^2+x^2)^2} ,\,
\frac{-q^2}{(q^2+x^2)^2} \right\} + diag \left\{
\frac{-2q^2}{(q^2+x^2)^2} ,\, 0 ,\, 0 ,\, 0 \right\}
\label{1-7}\end{equation} The first term on the right side
corresponds to the energy-momentum tensor of the electric (or
magnetic) field with a charge $q$; second - dust matter with
negative energy density.

\section{Linearization of the equations}
\label{s3}

Let us consider small spherical perturbations of matter and
metrics of the MTEB wormhole.

We introduce the notations:
\begin{equation}
8\pi\varepsilon \equiv \frac{-2q^2}{(q^2+x^2)^2}+f(x,t) ,\quad
e^\alpha \equiv (q^2+x^2) e^{\eta(x,t)} ,\quad \xi^2\equiv
(q^2+x^2) ,\quad 8\pi p=h f . \label{1-8}\end{equation} Here
${h(x)}$ --- arbitrary function, its physical meaning --- a square
of sound speed ${v_s^2}$ (in this fluid).

We introduce the dimensionless coordinates: namely we put ${q=1}$.

We write the equations (\ref{1-2}-\ref{1-5}) in the linear
approximation for small perturbations: ${\nu,\, \lambda,\, \eta,\,
f}$:
\begin{eqnarray}
f = \frac{\lambda - \eta - 3x\eta' + x\lambda'}{\xi^2} - \eta'' +
\frac{2\eta + \lambda}{\xi^4} ,
\label{1-9}\\
h f +\ddot\eta +\frac{\eta (1-x^2)}{\xi^4} +\frac{\lambda
x^2}{\xi^4} -\frac{x(\nu' + \eta')}{\xi^2} = 0 ,
\label{1-10}\\
h f = \frac{\nu''+\eta'' - \ddot\lambda - \ddot\eta}{2} +
\frac{2\eta - \lambda}{\xi^4} + \frac{x(\eta' - \lambda'/2 +
\nu'/2)}{\xi^2} ,
\label{1-11}\\
\dot\eta' +x(\dot\eta - \dot\lambda)/\xi^2 =0 . \label{1-12}
\end{eqnarray}
From (\ref{1-12}) we have:
\begin{eqnarray}
\eta - \lambda = F_1(x)- \frac{\xi^2}{x} \eta' \label{1-13}
\end{eqnarray}

If the pressure $p$ is isotropic, then from equations
(\ref{1-2}-\ref{1-5}) we can obtain two useful relations which
derived directly from the formula ${T^k_{i;k}=0}$ (consequence of
Bianchi identities):
\begin{equation}
\dot\lambda +2\dot\alpha = -2\dot\varepsilon /(p+\varepsilon) \,
,\quad \nu' =-2p'/(p+\varepsilon) \, .
\label{Bianki-1}\end{equation} Or in the linear approximation:
\begin{equation}
\dot\lambda +2\dot\eta = \xi^4\dot{f} \, ,\quad \nu' = \xi^4
(hf)'\, . \label{Bianki-2}\end{equation} From (\ref{Bianki-2}) we
have:
\begin{equation}
\xi^4 f = \lambda + 2\eta + F_2(x) \label{Bianki-3}\end{equation}
Expressing $\lambda$ from (\ref{1-13}), and taking into account
(\ref{Bianki-3}) from (\ref{1-9}) we have:
\begin{eqnarray}
F_2(x)=-\xi^2 (xF_1)' \label{1-14}\end{eqnarray} The arbitrary
functions $F_1(x)$ and $F_2(x)$ are determined by the choice of
initial conditions (at ${t=0}$) to small perturbations $\lambda$,
$f$ and $\nu$ (for given initial condition at ${\eta(x,t)}$).
Therefore, renaming at ${t=0}$ in (\ref{1-13}) and
(\ref{Bianki-3}) for small perturbations:
${\lambda\to\tilde\lambda - F_1}$ and ${f\to\tilde f + F_2/\xi^4}$
then we can put the function $F_1$ and $F_2$ equal to zero.

From (\ref{1-9}), (\ref{1-10}) and (\ref{1-13}) we have:
\begin{eqnarray}
f = 3\eta/\xi^4 + \eta'/(x\xi^2) = \left[\xi^3 \eta\right]'/(x\xi^5), \label{1-15-2}\\
\ddot\eta + h f-x\xi^2 (h f)' + \eta/\xi^4 =0 , \label{1-15-1}\\
\lambda = \frac{\xi}{x}\left[\xi \eta\right]' . \label{1-15-3}
\end{eqnarray}
From the equations (\ref{1-15-2}-\ref{1-15-1}) we obtain:
\begin{eqnarray}
\ddot\eta - h \eta'' + \eta'\left[ \frac{2h}{x\xi^2} - h'\right] +
\eta U(x) = 0 \label{1-16-0}\end{eqnarray} where
\begin{eqnarray}
U(x)\equiv \frac{12h x^2 + 3h - 3h' x\xi^2 + 1}{\xi^4}
\label{1-16-00}\end{eqnarray}

Equation (\ref{1-16-0}) at ${h>0}$ is a differential equation of
hyperbolic type. At infinity ${(x\to\pm\infty)}$ the potential
${U(x)}$ tends to zero and equation (\ref{1-16-0}) becomes an
equation for sound waves with the speed of sound ${v_s=\sqrt{h}}$.

\section{Stability investigation}
\label{s3-2}

We transform the equation (\ref{1-16-0}) to the canonical form. To
do this we change variables ${x\to z}$. Then ${\partial_x\to
\beta\partial_z}$, where ${\beta\equiv\frac{\partial z}{\partial
x}}$. Equation (\ref{1-16-0}) can be rewritten as:
\begin{eqnarray}
\ddot\eta - \left[h\beta^2\eta,_{zz} +h\beta'\eta,_z
-\beta\eta,_z\left( \frac{2h}{x\xi^2} - h'\right) \right] + \eta U
= 0 \label{1-16}\end{eqnarray} Now choose a function ${h(x)}$ and
${\beta (x)}$, so that in the square brackets expressions
(\ref{1-16}) was only
value of ${\eta,_ {zz}}$, ie ${h\beta^2 =1}$ and ${h\beta' - \beta [2h/(x\xi^2) - h'] =0}$. \\
From these conditions we obtain ${h=h_0 x^4/\xi^4}$, where
${h_0=const>0}$.

Then the hyperbolic equation (\ref{1-16-0}) can be rewritten in
the canonical form:
\begin{eqnarray}
\ddot\eta - \eta,_{zz} + \eta U(z) = 0 \label{1-16-1}
\end{eqnarray}
where
\begin{eqnarray}
U(z) = \frac{12h_0 x^6 - 9h_0 x^4 + (1+x^2)^2}{(1+x^2)^4}
\label{1-16-11}
\end{eqnarray}
 Here we took into account that now the variable $x$ depends
from the variable $z$ by the following equation:
\begin{eqnarray}
x^2-\sqrt{h_0}z x -1=0 \label{1-17-2}
\end{eqnarray}
and
\begin{eqnarray}
x^\pm(z)=\frac{\sqrt{h_0}z\pm\sqrt{h_0 z^2 +4}}{2} ,
\label{1-17-22}
\end{eqnarray}
Here the sign "$+$"\, corresponds to the ${x>0}$, and the sign
"$-$"\, --- respectively ${x<0}$. We first investigate the region
${x>0}$.

Asymptotics of ${z\to -\infty}$ corresponds to ${x\to
(\sqrt{h_0}|z|)^{-1}\to 0}$, and asymptotics ${z\to +\infty}$
corresponds to ${x\to (\sqrt{h_0}z)\to +\infty}$.

The solution of equation (\ref{1-16-1}) can be obtained by
separation of variables:
\begin{eqnarray}
\eta (z,t)=\sum\limits_{n=0}^{\infty} T_n(t) \Psi_n(z) , \label{1-17-1}\\
\frac{\ddot T_n}{T_n} = \frac{\Psi_{n,zz}}{\Psi_n} - U(z) = -w_n^2
. \label{1-17-3}
\end{eqnarray}
From (\ref{1-17-3}) we obtain: ${T_n=\exp(iw_n t)}$. Here, the
quantity ${w_n = const}$ has a physical sense to the harmonic
oscillation frequency with the number $n$, for the small
perturbation $\eta$.

From (\ref{1-17-3}) we obtain for each harmonic:
\begin{eqnarray}
\Psi_{n,zz} + w_n^2\Psi_n - U(z)\Psi_n = 0
\label{1-18}\end{eqnarray} The expression (\ref{1-18}) is a
stationary Schr\"odinger equation for the whole numerical axis
${(-\infty, +\infty)}$ with the potential ${U(z)}$.

\begin{figure*}
\subfigure[\label{R1_left} ${Side\quad view}$
]{\includegraphics[width=0.49\textwidth]{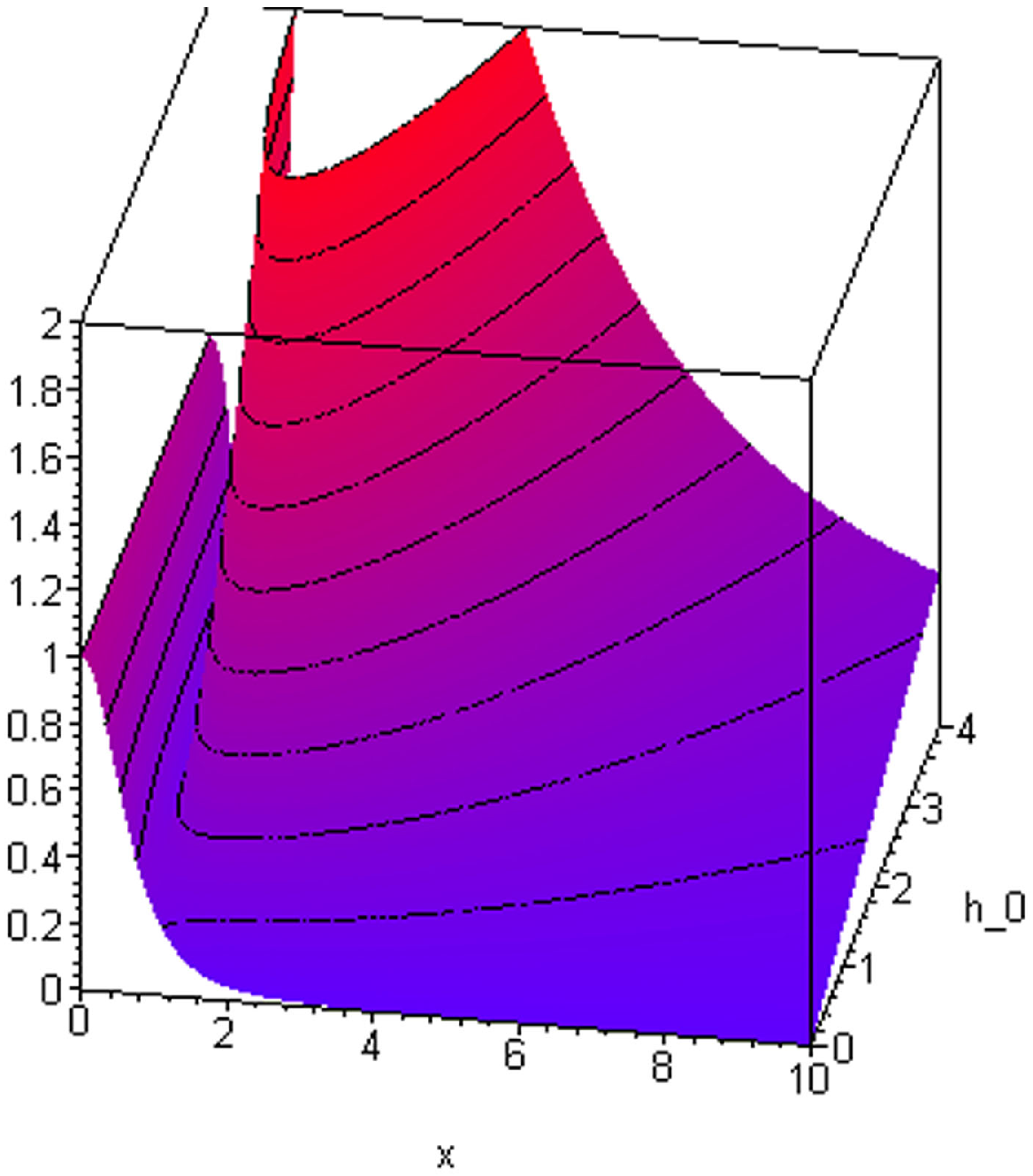}}
\subfigure[\label{R1_right} ${Top\quad view}$
]{\includegraphics[width=0.49\textwidth]{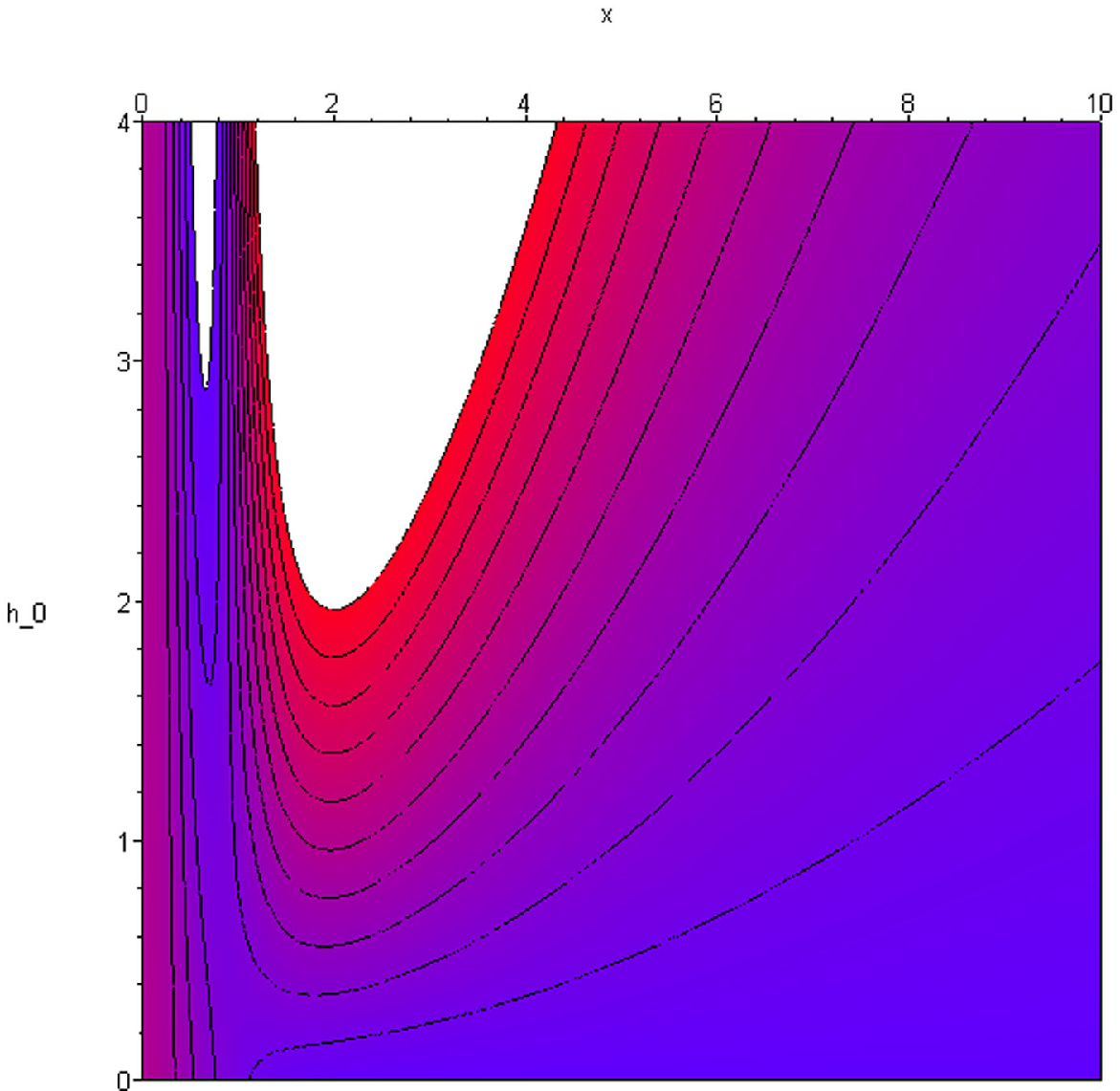}} \caption{{View
of the surface for the potential ${U(x,h_0)}$ for ${h(x)=h_0
x^4/\xi^4}$. This shows that region ${U>0}$ corresponds to the
range ${0<h_0<h_{00}\approx 2.8}$ for any values of $x$.}}
\label{R1}
\end{figure*}

As is well known (see.~\cite{Landau3}, \S18) energy levels
${E_n=(w_n \hbar)^2/2}$ in the Schr\"odinger operator spectrum
are always positive, if the effective potential ${U(z)}$ is
regular and non-negative (for all $z$) and ${U(z)\to 0}$ at
${z\to\infty}$. From (\ref{1-16-1}) follows that ${U(z)}$
satisfies these conditions at ${0<h_0<h_{00}\approx 2.8}$ ---
see~figure. However, the physical constraint on the parameter
${h_0}$ is the maximum speed of sound, which can not exceed the
speed of light. Since the square of the speed of sound is
${\frac{\partial p}{\partial\varepsilon}=h\le c^2}$, then there
must be ${h_{max}=h_0\le 1}$.

The result is ${w_n^2\ge 0}$, ie oscillation frequency must be real value.\\
The proof for region ${x<0}$ is analogous.

Thus we demonstrated the existence of the matter model in which
the function ${\eta (x,t)}$ is a nonincreasing.

Since each term of (\ref{1-17-1}) nonincreasing for time
evolution, there must be a non-increasing, and the derivative by
${x}$ from each member of this series (as derivative by $x$ does
not affect on the time component $T_n$, which responsible for the
time dependence). Therefore, we can make a statement that the
quantity ${\eta'(x,t)}$ well as nonincreasing (at ${0<h_0\le 1}$).
Therefore, the functions ${\nu (x,t)}$, ${f(x,t)}$ and ${\lambda
(x,t)}$ are nonincreasing (according to expressions
(\ref{Bianki-2}, \ref{1-15-2}-\ref{1-15-3})). Thus, the complete
solution is stable.

The case ${h = 0}$ should be considered individually: Eq.
(\ref{1-16}) at ${h=0}$ becomes an equation of parabolic type and
the variables ${(x,t)}$ are not separated, as decomposition in the
form (\ref{1-17-1}) becomes inapplicable. The solution of equation
(\ref{1-16-0}) with ${h = 0}$ is easily found in the form:
${\eta_{h=0}(x,t)=\eta_0 \exp \left( it/\xi^2 \right)}$. This
solution is also stable. However, the derivative of this solution
${\eta'_{h=0}=-2i t x \eta_{_{h=0}}/\xi^4}$ at ${x\ne 0}$ is not
asymptotically stable function, and has a linear instability with
respect to time. Consequently, the functions ${\lambda_{h=0}
(x,t)}$ and ${f_{h=0}(x,t)}$ at ${x\ne 0}$ are also linearly
unstable (see.~\cite{Novikov2009-1, Novikov2009-2}).

\section{Conclusions}
\label{s4}

We have demonstrated that in general relativity, it is possible to
construct a model of static and traversable wormhole, which will
be stable with respect to small spherical perturbations.

In this paper we do not perform investigation on the stability by
non-spherical perturbation types, but it is known
(see.~\cite{Bronnikov2008}), that the non-spherical perturbation
types, seems to be more stable than the spherical, because they
have a centrifugal (and other higher-multipole) barriers in the
effective potential for perturbations.

\section*{Acknowledgements}
\label{s5}

The authors are grateful to Kirill Bronnikov, Kip Thorne and Aaron
Zimmerman for discussions.

This work was supported in part by the Federal Program
``Scientific-Pedagogical Innovational Russia 2009-2011"\, and the
program by the presidium of RAS  "The origin, structure and
evolution of the universe 2011".

\end{document}